\documentclass[doublecol]{epl2}

\title{Heat transport in an open transverse-field Ising chain}

\author{Ke-Wei Sun \inst{1,2,3}, Chen Wang \inst{3}, and Qing-Hu Chen \inst{2,3,*}}

\institute{
 \inst{1} School of Science, Hangzhou Dianzi University,
Hangzhou 310018, P. R. China.\\
 \inst{2} Center for Statistical and
Theoretical Condensed Matter
Physics, Zhejiang Normal University, Jinhua 321004, P. R. China.   \\
 \inst{3} Department of Physics, Zhejiang University, Hangzhou 310027,
P. R. China.
}

\abstract{The heat conduction in an open transverse-field Ising
chain is studied by using quantization in the Fock space of
operators in the weak coupling regimes, i.e. the coupling is much
smaller than the transverse field. The non-equilibrium steady state
is obtained for large size systems coupled to Markovian baths at its
ends. The ballistic transport is observed in the uniform chain and
normal diffusion in the random-exchange chain. {In addition, the
ballistic-diffusive transition is found  at the intermediate
disorder regime.} The thermal conductivity $\kappa$ is also
calculated in the low and high temperature regimes. It is shown that
$\kappa$ decays as $\kappa\sim T^{-2}$ at high temperatures.}

\begin{document}
\maketitle

\section{Introduction}

Heat and spin transport behaviors of one-dimensional systems have
intensively been studied  in both classical and quantum mechanical
context for several
decades\cite{Yan1,Monasterio,Yan2,Zhang,Michel1,Zotos,Sologubenko,Saito,Mejia,Michel2,Michel3,Steinigeweg,Jung}.
Some classical nonlinear systems of interacting particles usually
show a diffusive behavior, which satisfy Fourier's law,
$J=-\kappa\nabla T$ ($\kappa$ is the thermal conductivity),
{relating} the macroscopic heat flux to the temperature
gradient\cite{Hu,Peierls,Dhar1}. In the quantum regime, it is also
an important issue that how normal energy or heat is transported
through a sample on a microscopic level. The ballistic behavior has
been proposed in the integrable quantum systems,  implying that the
current-current correlation functions typically do not decay to
zero\cite{Sologubenko}. The chaotic dynamics of a nonintegrable
system may  yield a normal diffusion. One natural question is how
the  heat transports through a quantum disordered system.

There exists two general theoretical approaches for a description of
non-equilibrium open quantum spin systems. One is the
non-equilibrium green's function method\cite{Kubo,Mori}. The other
is the quantum master equation \cite{Kryszewski,Wichterich}. In the
latter case,  various approximations schemes are employed, such as
Markov approximations, Born approximations, secular approximations,
and weak coupling approximations for the system-environment
coupling. Some effective dissipative equations of motion for reduced
density matrix $\hat{\rho_s}$ of the open systems are then derived.

By using the proper quantum master equation, some interesting and
important thermal properties have been found in the open spin-chain
systems by using the Monte Carlo wave function
method\cite{Michel1,Carlo,Lindblad,Breuer}, the matrix product
operator method\cite{Prosen3}, and fourth order Runge-Kutta
method\cite{Yan1,Yan2,Zhang}. However, the largest system sizes
reported in the {literature is} smaller than $N=20$, and the
convincing results for the properties and the type of the transport
are still lacking to date. In the meantime, some progresses in the
methodology have also been achieved, such as  the adaptive
time-dependent density matrix renormalization group, the numerically
exact diagonalization  and the quantization in the Fock space of
operators etc.\cite{Dubi,Langer,Steinigeweg1,Karevski,Prosen1},
which could be applied in the large system successfully. Therefore,
the studies on the large size system have become an essential and
intensive issues. Recently, Prosen\emph{et al.}\cite{Prosen1}
proposed a method to solve explicitly the Lindblad master equation
for an arbitrary quadratic system of $n$ fermions in terms of
diagonalization of a $4n\times4n$ matrix. This method has been
successfully applied to the far from equilibrium quantum phase
transition\cite{Prosen2,Pi} in one-dimensional XY spin chain for
size larger than $100$.

In this paper, we study the heat current properties of the well
known 1D transverse field Ising model with very large sizes. The
exchange couplings are considered to be both uniform and random. The
decisive conditions of the ballistic transport and the normal
transport are given. The paper is organized as follows: In Sec.II,
we describe the model and the scheme to solve the
quantum master equation in detail. Then the numerical results are
obtained in Sec.III. The conclusion is given in the
last section.

\section{MODEL AND METHOD}

The Hamiltonian for an open 1D transverse field Ising chain reads
\begin{eqnarray}
H&=&-\sum_{n=1}^{N}h\sigma_n^z-\sum_{n=1}^{N-1}J_n\sigma_n^x\sigma_{n+1}^x,
\end{eqnarray}
where $N$ is the number of spins, the operators $\sigma_n^x$ and
$\sigma_n^z$ are the Pauli matrices for the $n$th spin, $J_n$ is the
coupling parameter between the nearest-neighbor spins, and $h$ is
the transverse magnetic field. Here we take $h=1$. For the
disordered  system, $J_n$ is chosen to distribute  on a interval
$(0.05,0.15)$ uniformly, modeling the weak coupling condition. For
the uniform  system, we take $J_n=0.15$. {Considering the two
thermal baths at two ends and the coupling with the Ising chain, the
total Hamiltonian can be written as
\begin{eqnarray}
\mathcal{H}&=&H+H_L+H_R+H_{int},
\end{eqnarray}
where $H_{L(R)}=\sum_{k}^{L(R)}\hbar\omega_k a_k^{\dagger}a_k$ is
the left(right) phonon bath  with $a_k^{\dagger}$($a_k$)   the
phonon creation(annihilation) operator, and
$H_{int}=\sigma_{1(N)}^{\mp}\sum_k^{L(R)}(g_ka_k+g_k^*a_k^{\dagger})$
is the interaction between the chain and the baths. If the coupling
$g_k$ is weak, a quantum master equation for the system¡¯s evolution
can be obtained from our microscopic Hamiltonian model by using the
usual Born-Markov approximations and the secular
approximation\cite{Kryszewski}.}

The quantum master equation in the weak internal coupling
limit{($\{J_n\}\ll h$)} reads (we set $\hbar=1$)
\begin{eqnarray}
\frac{d\rho}{d t}=-i[H,\rho]+D_L(\rho)+ D_R(\rho),
\end{eqnarray}
where the dissipator $D_L$ refers to the left heat bath and $D_R$ to
the right one, depending on the full density operator $\rho$ of the
Hamiltonian (1). {Eq. (3)} can be rewritten as the Lindblad master
equation
\begin{eqnarray}
\frac{d{\rho}}{d
t}=\mathcal{\hat{L}}\rho:=-i[H,\rho]+\sum_{\mu}(2L_{\mu}\rho
L_{\mu}^{\dagger}-\{L_{\mu}^{\dagger}L_{\mu},\rho\}),
\end{eqnarray}
where $L_{\mu}$s are the Lindblad operators, representing couplings
to different baths. The weak bath coupling is taken into account
here. The simplest nontrivial bath operators acting only on the
first and the last spin are chosen ({$\mu=1,2,3$, and $4$})
\begin{eqnarray}
L_{1,2}=\sqrt{\Gamma_{1,2}^L}\sigma_1^{\mp}, \quad
L_{3,4}=\sqrt{\Gamma_{1,2}^R}\sigma_N^{\mp},
\end{eqnarray}
where $\sigma_m^{\mp}=(\sigma_m^x\pm i\sigma_m^y)/2$. Refer  to
Refs. \cite{Monasterio,Prosen1,Prosen2},  we have
$\Gamma_{1}^{L,R}=\pi \lambda_B[1+n_{L,R}(\omega)]
I(\omega)|_{\omega=2h}\equiv\lambda [1+n_{L,R}(2h)]$ and
$\Gamma_{2}^{L,R}=\pi\lambda_Bn_{L,R}(\omega)
I(\omega)|_{\omega=2h}\equiv\lambda n_{L,R}(2h)$. Here,
$n_{L,R}(\omega)=(e^{\omega/T_{L,R}}-1)^{-1}$ is the Bose-Einstein
distribution function ($k_B=1$),  $\lambda_B$ is the system-bath
coupling strength, and $I({\omega})$ denotes the spectral density of
an Ohmic bath that we choose. The Hamiltonian (1) is conveniently
expressed as a quadratic form $H=\underline{w}\cdot
\textbf{H}\underline{w}$ in terms of $2N$ Hermitian Majorana
operators
\begin{eqnarray}
w_{2n-1}=\sigma_n^x\prod_{n^{\prime}<n}\sigma_{n^{\prime}}^z,
\quad w_{2n}=\sigma_n^y\prod_{n^{\prime}<n}\sigma_{n^{\prime}}^z,
\end{eqnarray}
satisfying the anticommutation relation $\{w_n,w_m\}=2\delta_{n,m}$.
\textbf{H} is  an $2N\times2N$ antisymmetry Hermite matrix
($\textbf{H}^T=-\textbf{H}$). Based on the previous transformation,
we can rewrite Hamiltonian (1) in terms of Majorana fermions
\begin{eqnarray}
H&=&i\sum_{n=1}^{N}hw_{2n-1}w_{2n}+i\sum_{n=1}^{N-1}J_nw_{2n}w_{2n+1},
\end{eqnarray}
\begin{eqnarray}
L_{1,2}=\sqrt{\Gamma_{1,2}^L}(w_1\mp iw_2), \nonumber \\
L_{3,4}=-(-i)^N\sqrt{\Gamma_{1,2}^R}(w_{2N-1}\mp i w_{2N})W,
\end{eqnarray}
where $W=w_1w_2\cdots w_{2N}$ is a Casimir operator which commutes
with all the elements of the Clifford algebra generated by
$w_j${\cite{Prosen1}}. Note that  $W^2=1$, so it does not affect the
quadratical system. For convenience, we take $W=1$.

Then we construct $4^N$ dimensional Pauli algebra with a Fock space
of operators describing $2N$ adjoint fermions (a-fermions), with an
orthonormal canonical basis
$|P_{\underline{\alpha}}\rangle=|w_1^{\alpha_1}w_2^{\alpha_2}\cdots
w_{2N}^{\alpha_{2N}}\rangle$, $\alpha_j\in\{0,1\}$. With  the
definition
$\hat{c}_j|P_{\underline{\alpha}}\rangle=\delta_{\alpha_{j,1}}|w_jP_{\underline{\alpha}}\rangle$,
the quantum Liouvillean {(4)} becomes bilinear
$\mathcal{\hat{L}}=\underline{\hat{a}}\cdot\textbf{A}\underline{\hat{a}}+const
\bf1$ in Hermitian maps
$\hat{a}_{2j-1}=\frac{1}{\sqrt{2}}(\hat{c}_j+\hat{c}_j^{\dagger})$,
$\hat{a}_{2j}=\frac{1}{\sqrt{2}}(\hat{c}_j-\hat{c}_j^{\dagger})$,
obeying $\{\hat{a}_p,\hat{a}_q\}=\delta_{p,q}$. The $4N\times4N$
matrix $\textbf{A}$ can be expressed in a block tridiagonal form in
terms of $4\times4$ matrices as
\begin{equation}
\textbf{A}=
\left(
\begin{array}{ c c c c l r }
\textbf{B}_L-2h\textbf{R} & \textbf{R}_1 & \textbf{0} & \dots &
\textbf{0} \\
-\textbf{R}_1^T & -2h\textbf{R} & \textbf{R}_2 & \ddots & \textbf{0} \\
\textbf{0} & -\textbf{R}_2^T & -2h\textbf{R} &  & \vdots \\
\vdots & \ddots &  & \ddots & \textbf{R}_{N-1}\\
\textbf{0} & \textbf{0} & \dots & -\textbf{R}_{N-1}^T &
\textbf{B}_R-2h\textbf{R}
\end{array}
\right),
\end{equation}
where
\begin{equation}
\textbf{B}_{L,R}=
\left(
\begin{array}{ c c c c l r }
0 & 2i\Gamma_+^{L,R} & -2i\Gamma_-^{L,R} & 2\Gamma_-^{L,R}\\
-2i\Gamma_+^{L,R} & 0 & 2\Gamma_-^{L,R} & 2i\Gamma_-^{L,R} \\
2i\Gamma_-^{L,R} & -2\Gamma_-^{L,R} & 0 & 2i\Gamma_+^{L,R} \\
- 2\Gamma_-^{L,R} & -2i\Gamma_-^{L,R} & -2i\Gamma_+^{L,R} & 0
\end{array}
\right)
\end{equation}
with $\Gamma_{\pm}^{L,R}=\Gamma_2^{L,R}\pm\Gamma_1^{L,R}$ and
\begin{equation}
\textbf{R}=
\left(
\begin{array}{ c c c c l r }
0 & 0 & -1 & 0\\
0 & 0 & 0 & -1 \\
1 & 0 & 0 & 0 \\
0 & 1 & 0 & 0
\end{array}
\right),
\textbf{R}_m= \left(
\begin{array}{ c c c c l r }
0 & 0 & 0 & 0\\
0 & 0 & 0 & 0 \\
2J_m & 0 & 0 & 0 \\
0 & 2J_m & 0 & 0
\end{array}
\right).
\end{equation}
The eigenvalues of $4N\times4N$ antisymmetric $\textbf{A}$ called
rapidities can be list in the form of the pairs
$\beta_1,-\beta_1,\beta_2,-\beta_2,\dots,\beta_{2N},-\beta_{2N}$,
Re$\beta_j\ge0$. The corresponding eigenvectors $\underline{v}_p$ (
$p=1,\dots,4N$ ) can be  defined by
$\textbf{A}\underline{v}_{2j-1}=\beta_j\underline{v}_{2j-1}$ and
$\textbf{A}\underline{v}_{2j}=-\beta_j\underline{v}_{2j}$.
$\underline{v}$ can be normalized by using
$\underline{v}_{2j-1}\cdot\underline{v}_{2j}=1$ and
$\underline{v}_{p}\cdot\underline{v}_{q}=0$ otherwise, which  can be
used to calculate any quadratic physical observable in the
non-equilibrium steady state (NESS). The expectation value is given
by
\begin{eqnarray}
\langle w_jw_k\rangle_{NESS}&=&\delta_{j,k} +\frac 1
2\sum_{n=1}^{2N}(v_{2n,2j-1}-iv_{2n,2j})\nonumber\\
&&\times(v_{2n-1,2k-1}-iv_{2n-1,2k}).
\end{eqnarray}

\section{NUMERICAL RESULTS}
\begin{figure}[tbp]
\includegraphics[scale=0.8]{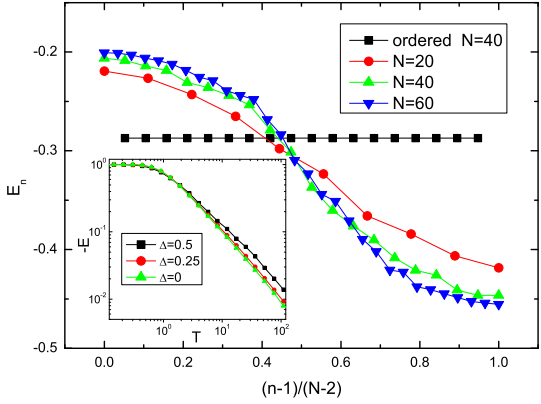}
\caption{(color online) Energy profiles $E_n$ (odd sites) for a
random chain with different sizes $N$.  $J_n\in(0.05,0.15)$,
$T_L=5$, $T_R=2$. For comparison, the energy profile $E_n$ of an
ordered chain with $J_n=J=0.1$ is also listed. The inset denotes the
local energy $E_2$ versus average temperature $T$ for the disordered
system ($N=4$) with different temperature difference $\Delta$. All
results are averaged over 200 realizations of disorder.}
\label{Figure1}
\end{figure}

First, to show the correctness of this method, we numerically
calculate the local energy $E_n$. The local energy density
operator reads
\begin{eqnarray}
H_n=iJ_nw_{2n}w_{2n+1}+i\frac h 2
(w_{2n-1}w_{2n}+w_{2n+1}w_{2n+2}),
\end{eqnarray}
and local energy density is defined as $E_n=\langle
H_n\rangle_{NESS}$. {The temperatures of the left and right baths
are $T_L=T(1+\Delta)$ and $T_R=T(1-\Delta)$, where $\Delta$ is the
dimensionless temperature difference and $T$ is the average
temperature.} We set the coupling constants uniformly distribute on
the interval $J_n\in(0.05,0.15)$, which are the same  as those in
Ref. \cite{Yan1}. $\lambda$ is taken as 0.005, which is equivalent
to the parameter $\alpha=0.01$ in the  Ref. \cite{Yan1}. {The
numerical results for local energy $E_n$ with different sizes are
presented in Fig. 1. Here we only plot the local energy on the odd
sites for the similar results on the even sites. Note that there is
a clear intersection of the energy profiles $E_n$ (disordered) for
different sizes at the central part of the chain. And the inset
shows $E_2$ versus $T$ for $N=4$. The local energy decreases with
the increase of   $\Delta$. When $\Delta=0$, the results can be
checked with the canonical one\cite{Yan1}.} It is interesting that
the present local energy is nearly the same as those in Ref.
\cite{Yan1}. It should be pointed out here that the method is not
suited to the strong coupling case, since the Lindblad equation is
only valid for small $\{J_n\}$.

\begin{figure}[tbp]
\includegraphics[scale=0.5]{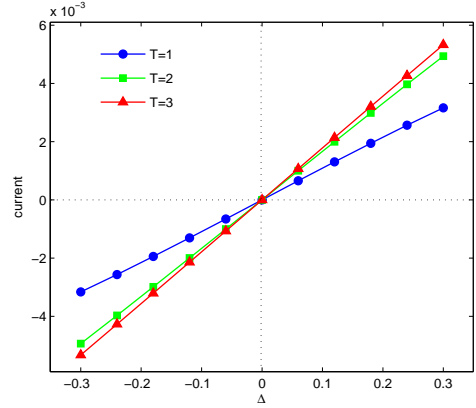}
\caption{(color online) Heat current vs temperature difference for
the uniform system. The average temperature $T$ changes from 1 to 3.
The other parameters are $h=1$, $J_n=0.15$, $\lambda=0.005$, and
$N=100$.} \label{Figure2}
\end{figure}
\begin{figure}[tbp]
\includegraphics[scale=0.5]{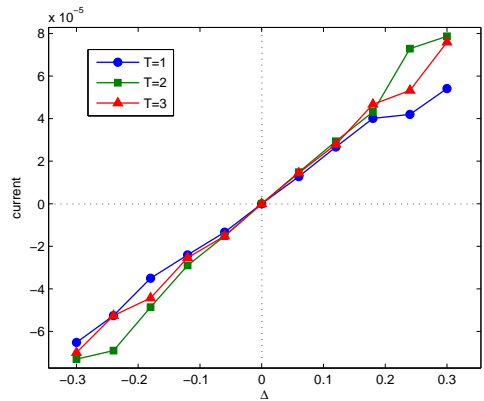}
\caption{(color online) Heat current vs temperature difference for
the disordered  system. The range of $T$ is set from 1 to 3. The
other parameters are $h=1$, $J_n\in(0.05,0.15)$, $\lambda=0.005$,
and $N=100$. The results  are averaged over 100 realizations of
disorder.} \label{Figure3}
\end{figure}
Then we evaluate the  heat current in the spin chain in NESS
\begin{eqnarray}
Q_n&=&i[H_n,H_{n+1}]\nonumber \\
&=&i[-J_nhw_{2n}w_{2n+2}-hJ_{n+1}w_{2n+1}w_{2n+3}].
\end{eqnarray}
The heat current as a function of the temperature difference for
$N=100$ chain is presented in Fig. 2 for the uniform case  and Fig.
3 for the disordered case with the mean temperature ranging from
$T=1$ to $3$. According to {Eq. (5)}, the high bath temperature can
enhance the coupling strength between the baths and the spin chain,
which facilitates the heat transfer. So $\kappa$ increases with the
augmentation of the average temperature $T$ in the given parameters
regime, as shown in Fig. 2. For the disordered case, $\kappa$
changes little in small $\Delta$ regimes ($-0.1<\Delta<0.1$), as
shown in Fig. 3.  {From the order of magnitude of the heat current,
one can see that the current is easily blocked for the disordered
case.}
\begin{figure}[tbp]
\includegraphics[scale=0.5]{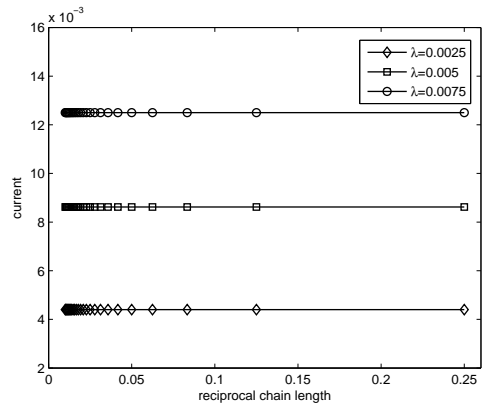}
\caption{Heat current versus reciprocal chain length for different
bath coupling strength $\lambda$ for the uniform system. The
parameters are $J_n=0.15$, $h=1$, $\beta_L=1/T_L=0.25$ and
$\beta_R=1/T_R=0.75$.} \label{Figure4}
\end{figure}
\begin{figure}[tbp]
\includegraphics[scale=0.75]{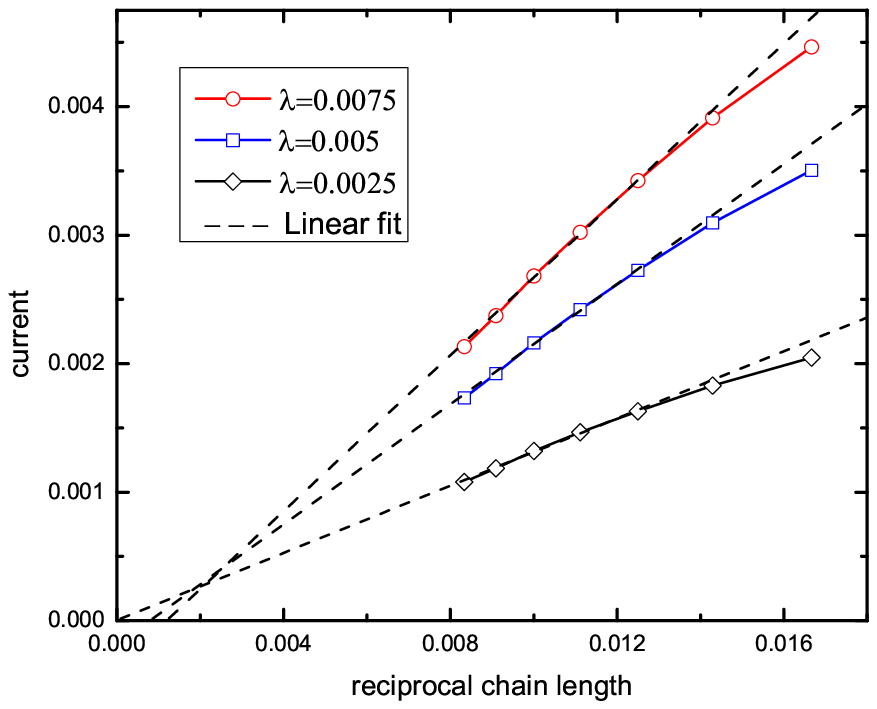}
\caption{(color online) Heat current  versus reciprocal chain length
for the disordered system. System parameters: $J_n\in(0.05,0.15)$,
$h=1$, $\beta_L=1/T_L=0.25$ and $\beta_R=1/T_R=0.75$ . The sizes of
the system range from $60$ to $120$ every other $10$ sites. The
results  are averaged over $2000$ realizations of disorder. }
\label{Figure5}
\end{figure}

We turn to discuss the classification of heat transport properties.
Note that a finite current within an infinite system demonstrates
ballistic transport behavior. In the previous studies,  the
ballistic behavior is observed in the integrable system, and the
diffusive transport occurs for the disordered system. But these
conclusions were built on the numerical simulations on  small
systems\cite{Yan1,Yan2,Michel1,Michel2,Michel3}.

\begin{figure}[tbp]
\includegraphics[scale=0.8]{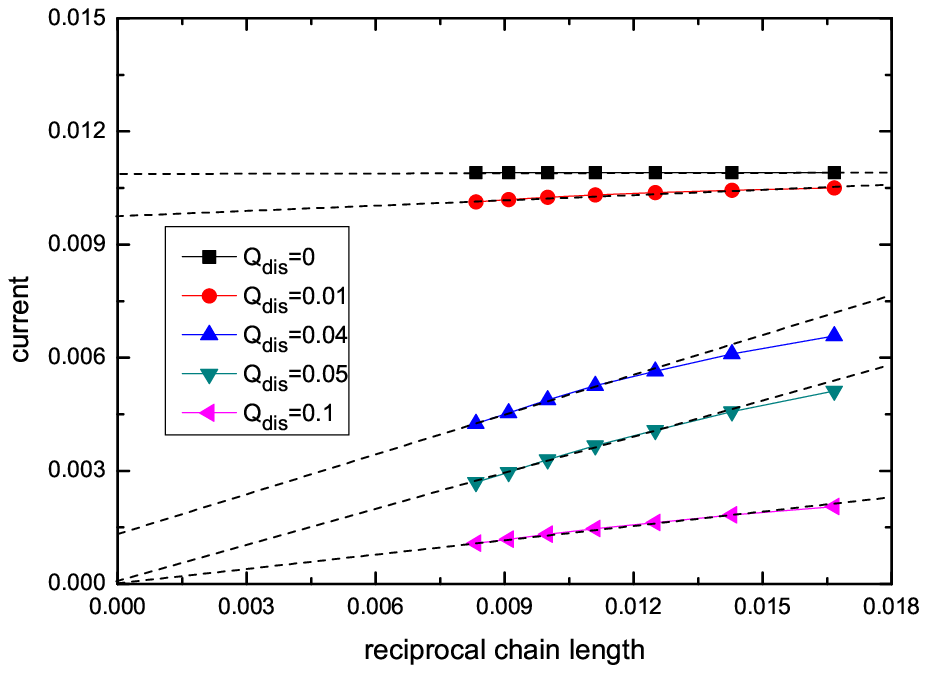}
\caption{(color online) Heat current versus reciprocal chain length
for different disorder strengthes. System parameters are $h=1$,
$\lambda=0.0025$, $\beta_L=1/T_L=0.25$, and $\beta_R=1/T_R=0.75$.
The results are averaged over $2000$ realizations of disorder. }
\label{Figure6}
\end{figure}
\begin{figure}[tbp]
\includegraphics[scale=0.8]{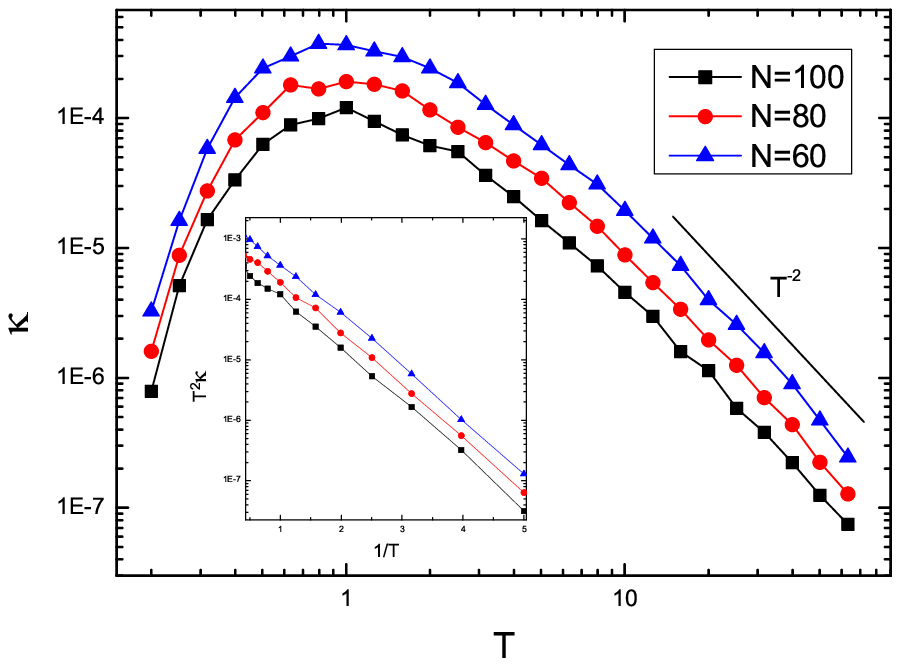}
\caption{Thermal conductivity $\kappa$ versus $T$ for the disordered
system with $J_n\in(0.05,0.15)$, $h=1$ and different sizes
$N=60,80,100$. The bath temperatures are $T_L=T(1+\Delta)$ and
$T_R=T(1-\Delta)$ ($\Delta=0.02$). The results  are averaged over
$200$ realizations of disorder. The inset shows the low temperature
property of thermal conductivity.} \label{Figure7}
\end{figure}
To classify transport properties more convincingly, we simulate the
system with the size up to $N>100$ for both cases. The numerical
results are plotted in Figs. 4 and  5.  Note that the ballistic
transport occurs at the uniform case, as shown in Fig. 4.  The heat
current increases with the bath coupling strength $\lambda$.
Different from the observation in a Heisenberg chain\cite{Michel1},
the present heat current is  only sensitive to the parameter
$\lambda$ and independent of the system size, {which is just the
characteristics of the ballistic transport only emerging in a
quantum  system. The known Fourier's law is obviously invalid.} For
the disordered case, the current is gradually reduced with
increasing the system size, demonstrating the diffusive transport,
as exhibited in Fig. 5. {The similar results are also obtained in
the mass-disordered harmonic crystals\cite{Dhar}.} For the present
system size $N=100$,  $1/N$ scaling behavior of the heat current is
observed for $\lambda=0.0025$, demonstrating that the Fourier's law
holds in this case. So we provide an evidence of the macroscopic
heat transport in the quantum disordered system. With the increase
of the coupling parameter $\lambda$, the linear fit gradually
deviates from the origin of the coordinate, {implying that the
finite-size effect is more obvious for large coupling parameters,
thus the data for large  systems would be essential to get the
correct scaling for  strong coupling.}

{At this stage, it should be expected that  the ballistic-diffusive
transition might occur at the intermediate disorder regime. The
disorder strength is introduced in the coupling parameter
$J_n=0.1+Q_{dis}\varepsilon $, where $\varepsilon $ distributes
uniformly on the interval $(-0.5,0.5)$. We plot the heat current vs
reciprocal chain length for different disorder strengthes in Fig. 6.
The linear fit of the data shows that the transition occurs at
$Q_{dis}=0.04\sim0.05$.}

Finally, we obtain the fully thermal conductivity $\kappa$ as a
function of temperature in the different large sizes for the
disordered case. In the high temperature regime, it is observed that
$\kappa$ decays slightly faster than $T^{-2}$, as shown in {Fig. 7}.
$\kappa$ reaches a maximum value around $T=1.0$ {with different
sizes $N$}. The thermal conductivity decreases with increasing
system size. The inset presents the relation
$ln(T^2\kappa)\propto1/T$ in the low temperature regime. The thermal
conductivity $\kappa$ and the specific heat $c_v$ of a single spin
have a similar temperature dependence. These observations are
consistent with the analytic and numerical results of Ref. 1 based
on a smaller system.

\section{Conclusions}

In this paper, we have studied the heat transport behaviors of an
open Ising chain within master equation formalism by using
quantization in the Fock space of operators. The classification of
the transport properties is performed  in large size system (over
$100$ sites). Compared with the Monte Carlo wave-function method,
the precision has been considerably improved in the present
approach. We confirm the ballistic current in the uniform system
which is integrable. The normal transport is clearly observed in the
disordered system. The bulk conductivity decreases with the increase
of the system size. The heat current exhibits a diffusive behavior
above a critical interaction strength, which follows Fourier's law
in the normal transport. {Moreover, the ballistic-diffusive
transition occurs at the intermediate disorder regime.} It is also
observed that the thermal conductivity $\kappa$ has the similar
temperature dependence as the specific heat $c_v$ in the weak
coupling regime.

\acknowledgments

The authors acknowledge useful discussions with B. Li. This work was
supported by National Natural Science Foundation of China, PCSIRT
(Grant No. IRT0754) in University in China, National Basic Research
Program of China (Grant No. 2009CB929104), Zhejiang Provincial
Natural Science Foundation under Grant No. Z7080203, and Program for
Innovative Research Team in Zhejiang Normal University.

$*$ Corresponding author. Email:qhchen@zju.edu.cn


\begin{thebibliography}{100}

\bibitem{Yan1} Y. H. Yan, C. Q. Wu, G. Casati, T. Prosen, and B. W. Li, Phys. Rev. B \textbf{77},
(2008) 172411.
\bibitem{Monasterio} C. Mejia-Monasterio and H. Wichterich, Eur. Phys. J. Special Topics \textbf{151}, PP. 113-125
(2007).
\bibitem{Yan2} Y. H. Yan, C. Q. Wu, and B. W. Li, Phys. Rev. B \textbf{79}, (2009) 014207.
\bibitem{Zhang} L. F. Zhang, Y. H. Yan, C. Q. Wu, J. S. Wang, and B. W. Li, Phys. Rev. B \textbf{80}, (2009) 172301.
\bibitem{Michel1} M. Michel and O. Hess, Phys. Rev. B \textbf{77}, (2008) 104303.
\bibitem{Zotos} X. Zotos, F. Naef, and P. Prelov$\breve{s}$k, Phys. Rev. B \textbf{55}, (1997) 11029.
\bibitem{Sologubenko} A. V. Sologubenko, E. Felder, K. Giann$\grave{O}$, H. R. Ott, A. Vietkine,
and A. Revcolevschi, Phys. Rev. B \textbf{62}, (2000) R6108; A. V.
Sologubenko, K. Giann$\grave{O}$, H. R. Ott, A. Vietkine, and A.
Revcolevschi, ibid. \textbf{64}, (2001) 054412.
\bibitem{Saito} K. Saito, Europhys. Lett. \textbf{61}, (2003) 34.
\bibitem{Mejia} C. Mejia-Monasterio, T. Prosen, and G. Casati, Europhys. Lett.
\textbf{72}, (2005) 520.
\bibitem{Michel2} M. Michel, M. Hartmann, J. Gemmer, and G. Mahler, Eur. Phys.
J. B \textbf{34}, (2003) 325.
\bibitem{Michel3} M. Michel, G. Mahler, and J. Gemmer, Phys. Rev. Lett. \textbf{95},
(2005) 180602.
\bibitem{Steinigeweg} R. Steinigeweg, J. Gemmer, and M. Michel, Europhys. Lett. \textbf{75},
(2006) 406.
\bibitem{Jung} P. Jung, R. W. Helmes, and A. Rosch, Phys. Rev. Lett.
\textbf{96}, (2006) 067202.
\bibitem{Hu} B. Hu, B. Li, and H. Zhao, Phys. Rev. E \textbf{57}, (1998) 2992; B. Hu,
B. Li, and H. Zhao, ibid. \textbf{61}, (2000) 3828; K. Aoki and D.
Kusnezov, Phys. Lett. A \textbf{265}, (2000) 250.
\bibitem{Peierls} R. E. Peierls, Quantum Theory of Solids (Oxford University Press, London, 1955).
{\bibitem{Dhar1} K. Saito and A. Dhar, Phys. Rev. Lett.
\textbf{104}, (2010) 040601.}
\bibitem{Kubo} R. Kubo, J. Phys. Soc. Jpn. \textbf{12}, (1957) 570.
\bibitem{Mori} H. Mori, Phys. Rev. \textbf{115}, (1959) 298.
\bibitem{Kryszewski} S. Kryszewski and J. Czechowska-Kryszk, arXiv:quant-ph/0801.1757v1 (2008).
\bibitem{Wichterich} H. Wichterich, M. J. Henrich, H. P. Breuer, J. Gemmer,
and M. Michel, Phys. Rev. E \textbf{76}, (2007) 031115.
\bibitem{Carlo} G. G. Carlo, G. Benenti, and G. Casati, Phys.
Rev. Lett. \textbf{91}, (2003) 257903.
\bibitem{Lindblad}G. Lindblad, Commun. Math. Phys. \textbf{48}, (1976) 119.
\bibitem{Breuer} H.-P. Breuer and F. Petruccione, The theory of open
quantum systems, (Oxford University Press, London, 2002).
\bibitem{Prosen3}T. Prosen and Marko $\breve{Z}$nidari$\breve{c}$, J. Stat. Mech. (2009) P02035.
\bibitem{Dubi} Y. Dubi and M. D. Ventra, Phys. Rev. B \textbf{79},
(2009) 115415.
\bibitem{Langer} S. Langer, F. Heidrich-Meisner, J. Gemmer, I. P. McCulloch, and U. Schollw\"{o}ck,
Phys. Rev. B \textbf{79}, (2009) 214409.
\bibitem{Steinigeweg1} R. Steinigeweg and J. Gemmer, Phys. Rev. B \textbf{80},
(2009) 184402.
\bibitem{Karevski} D. Karevski and T. Platini, Phys. Rev. Lett. \textbf{102},
(2009) 207207.
\bibitem{Prosen1} T. Prosen, New J. Phys. \textbf{10}, (2008) 043026.
\bibitem{Prosen2} T. Prosen and I. Pi$\breve{z}$orn, Phys. Rev. Lett. \textbf{101},
(2008) 105701.
\bibitem{Pi} I. Pi$\breve{z}$orn and T. Prosen, Phys. Rev. B \textbf{79},
(2009) 184416. {\bibitem{Dhar} A. Chaudhuri, A. Kundu, D. Roy, A.
Dhar, J. L. Lebowitz, and H. Spohn, Phys. Rev. B \textbf{81}, (2010)
064301.}
\end{thebibliography}
\end{document}